\documentclass[apj]{emulateapj}

\usepackage{graphicx}
\usepackage{amssymb}
\usepackage{amsmath}
\usepackage{natbib}
\usepackage{color}
\usepackage[dvipsnames]{xcolor}
\usepackage{multirow}
\usepackage[breaklinks,colorlinks,citecolor=blue,linkcolor=magenta]{hyperref}
\shortauthors{Zhou et al.}

\newcommand{\ima}{\texttt{ima} files }
\newcommand{\flt}{\texttt{flt} files }

\newcommand{\tinytim}{\textit{Tiny Tim}}
\newcommand{\bpic}{$\beta$ Pic}
\newcommand{\vsini}{$v\sin i$}
\newcommand{\mjup}{M$_{\mbox{Jup}}$}
\newcommand{\revise}[1]{\textbf{{\color{cyan}{#1}}}}
\renewcommand{\revise}{}
\newcommand{\reviseTwo}[1]{\textbf{{\color{cyan}{#1}}}}
\renewcommand{\reviseTwo}{}
\newcommand{\period}{$10.7^{+1.2}_{-0.6}$}
\newcommand{\Jperiod}{$11.1^{+2.0}_{{-1.2}}$}
\newcommand{\Jperiodd}{$11.9^{+3.4}_{-1.8}$}
\newcommand{\Jamp}{$1.36\pm0.23 \%$}
\newcommand{\Hperiod}{$9.3^{+2.0}_{{-0.8}}$}
\newcommand{\Hperiodd}{$10.7^{+4.4}_{{-1.4}}$}
\newcommand{\Hamp}{$0.78\pm0.22 \%$}
\begin{document}

\title{Discovery of Rotational Modulations in the Planetary-mass
  Companion 2M1207\lowercase{b}: Intermediate Rotation Period and Heterogeneous Clouds in a Low
  Gravity Atmosphere}
\shorttitle{Variability of 2M1207b}
\author{Yifan Zhou\altaffilmark{1}, D\'aniel Apai\altaffilmark{1,2,3},
  Glenn H Schneider\altaffilmark{1},  Mark S. Marley\altaffilmark{4},
  Adam P. Showman\altaffilmark{2}}

\altaffiltext{1}{Department of Astronomy/Steward Observatory, The
  University of Arizona, 933 N. Cherry Ave., Tucson, AZ, 85721, USA,
  \href{mailto:yifzhou@email.arizona.edu}{yifzhou@email.arizona.edu}
}
\altaffiltext{2}{Department of Planetary Science/Lunar and Planetary Laboratory, The University of
  Arizona, 1640 E. University Blvd., Tucson, AZ 85718, USA}
\altaffiltext{3}{Earths in Other Solar Systems Team, NASA Nexus for
  Exoplanet System Science}
\altaffiltext{4}{NASA Ames Research Center, Naval Air Station,
  Moffett Field,Mountain View, CA 94035, USA}

\begin{abstract}
  Rotational modulations of brown dwarfs have recently provided
  powerful constraints on the properties of ultra-cool atmospheres,
  including longitudinal and vertical cloud structures and cloud
  evolution. Furthermore, periodic light curves directly probe the
  rotational periods of ultra-cool objects.  We present here, for the
  first time, time-resolved high-precision photometric measurements of
  a planetary-mass companion, 2M1207b.  We observed the binary
  system with HST/WFC3 in two bands and with two spacecraft roll
  angles. Using point spread function-based photometry, we reach a
  nearly photon-noise limited accuracy for both the primary and the
  secondary. While the primary is consistent with a flat light curve,
  the secondary shows modulations that are clearly detected in the
  combined light curve as well as in different subsets of the data.
  The amplitudes are 1.36\% in the F125W and 0.78\% in the F160W
  filters, respectively.  \reviseTwo{By fitting sine waves to the
  light curves,} we find a consistent period of
\period{} hours and
  similar phases in both bands. The J- and H-band amplitude ratio of
  2M1207b is very similar to a field brown dwarf that has identical
  spectral type but different J-H color.  Importantly, our study also
  measures, for the first time, the rotation period for a directly 
  imaged \reviseTwo{extra-solar planetary-mass companion.}
\end{abstract}

\maketitle
\section{Introduction}

\begin{figure*}
  \centering
  \plottwo{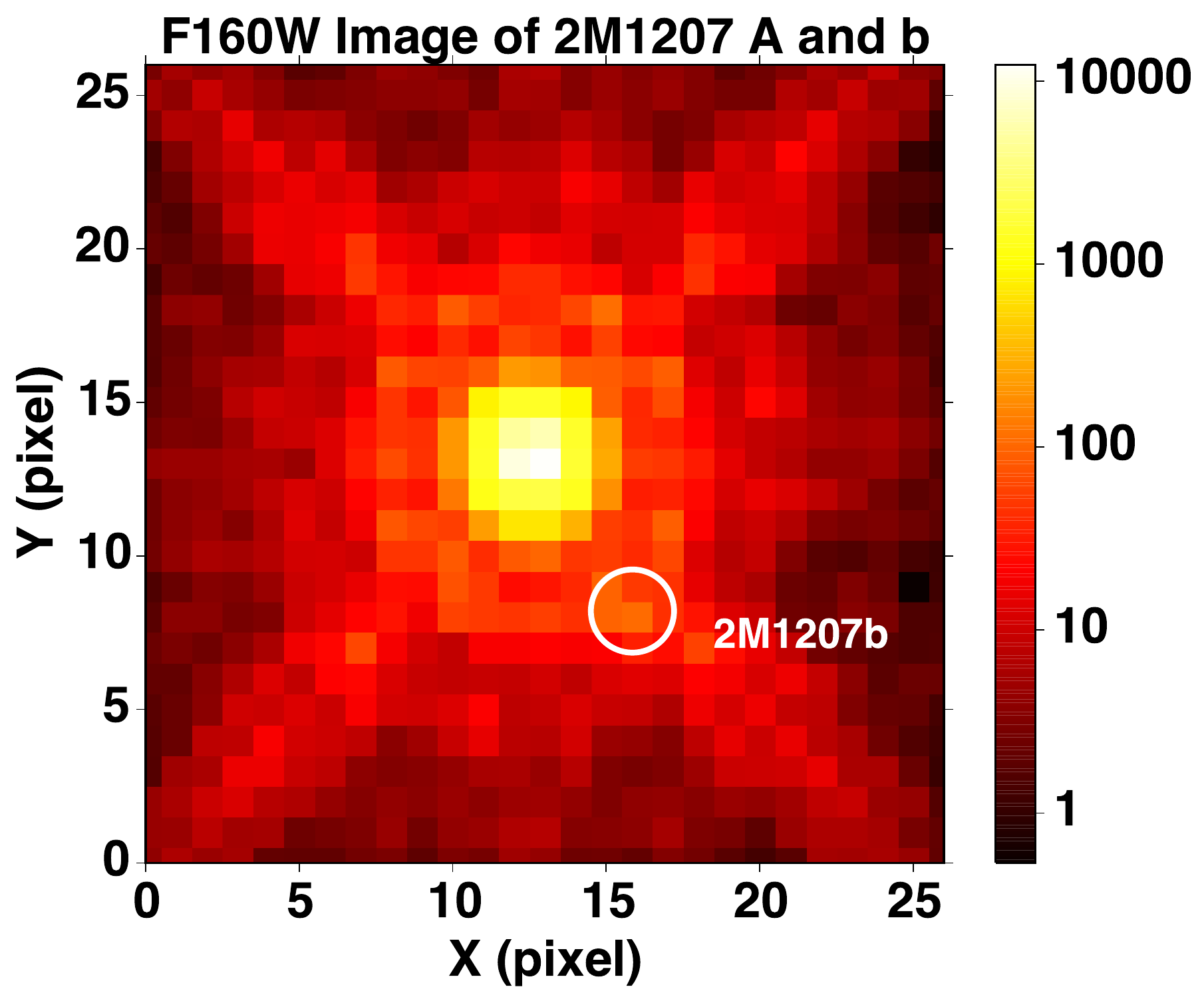}{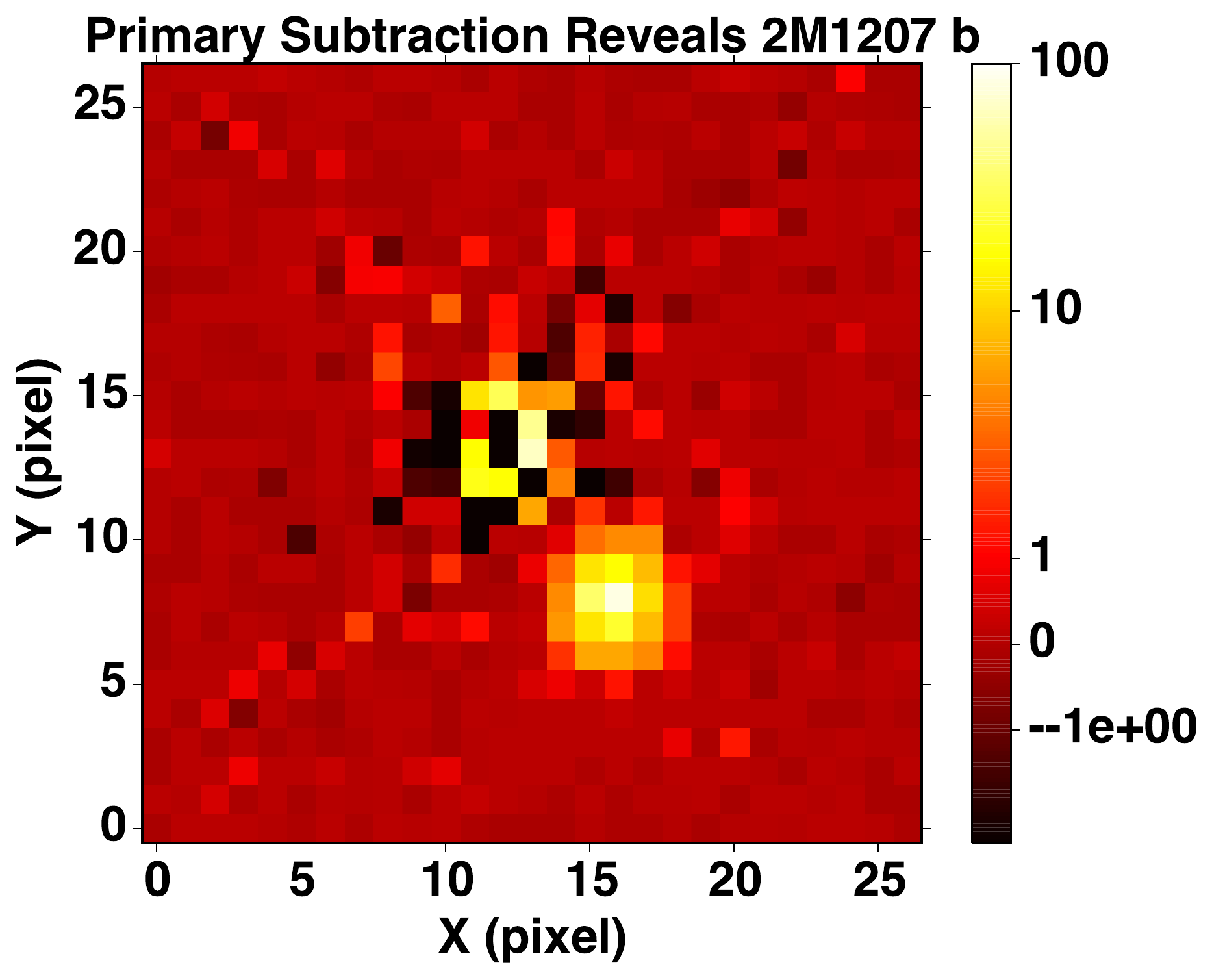}\\
  \includegraphics[width=0.32\textwidth]{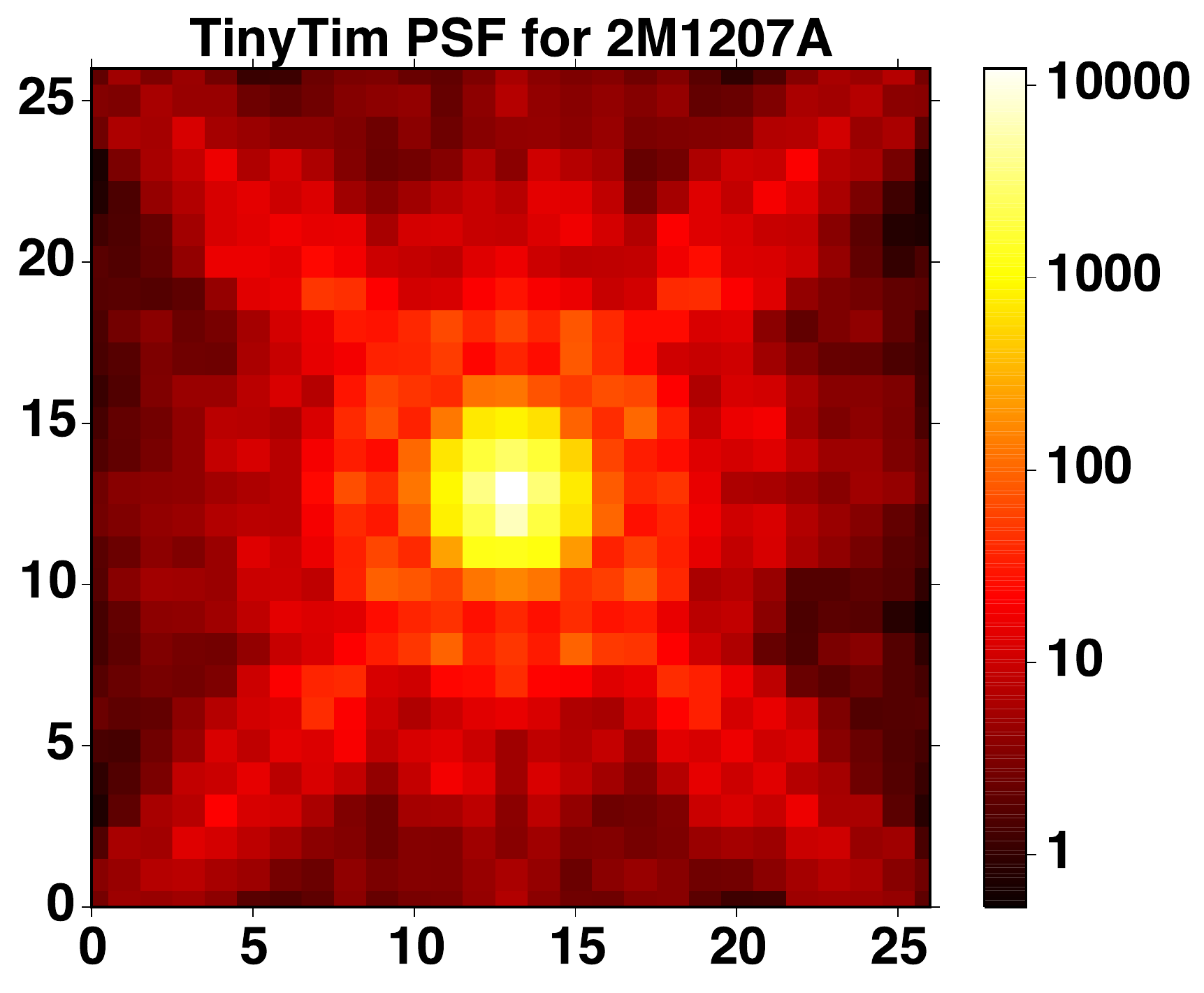}
  \includegraphics[width=0.32\textwidth]{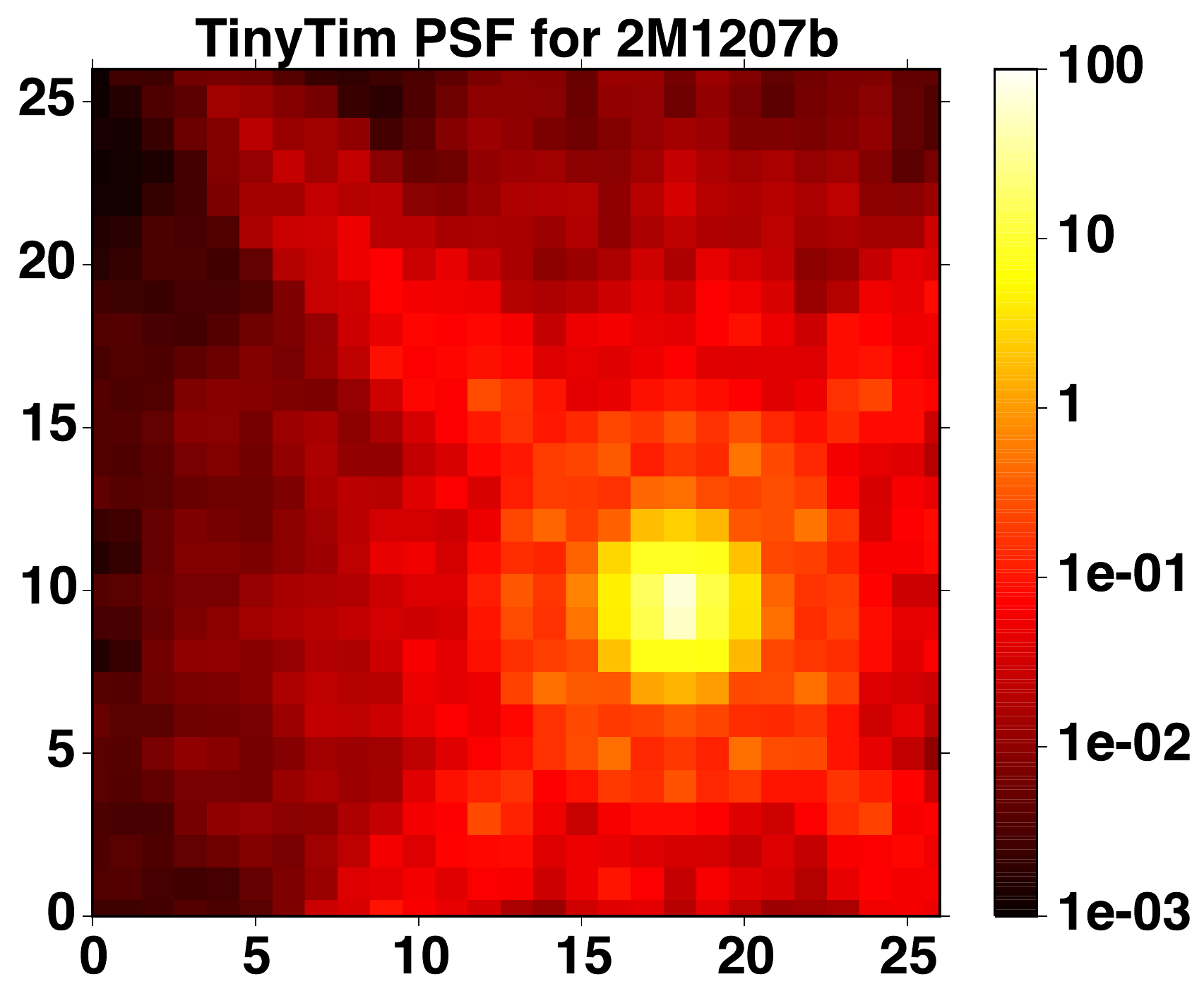}
  \includegraphics[width=0.32\textwidth]{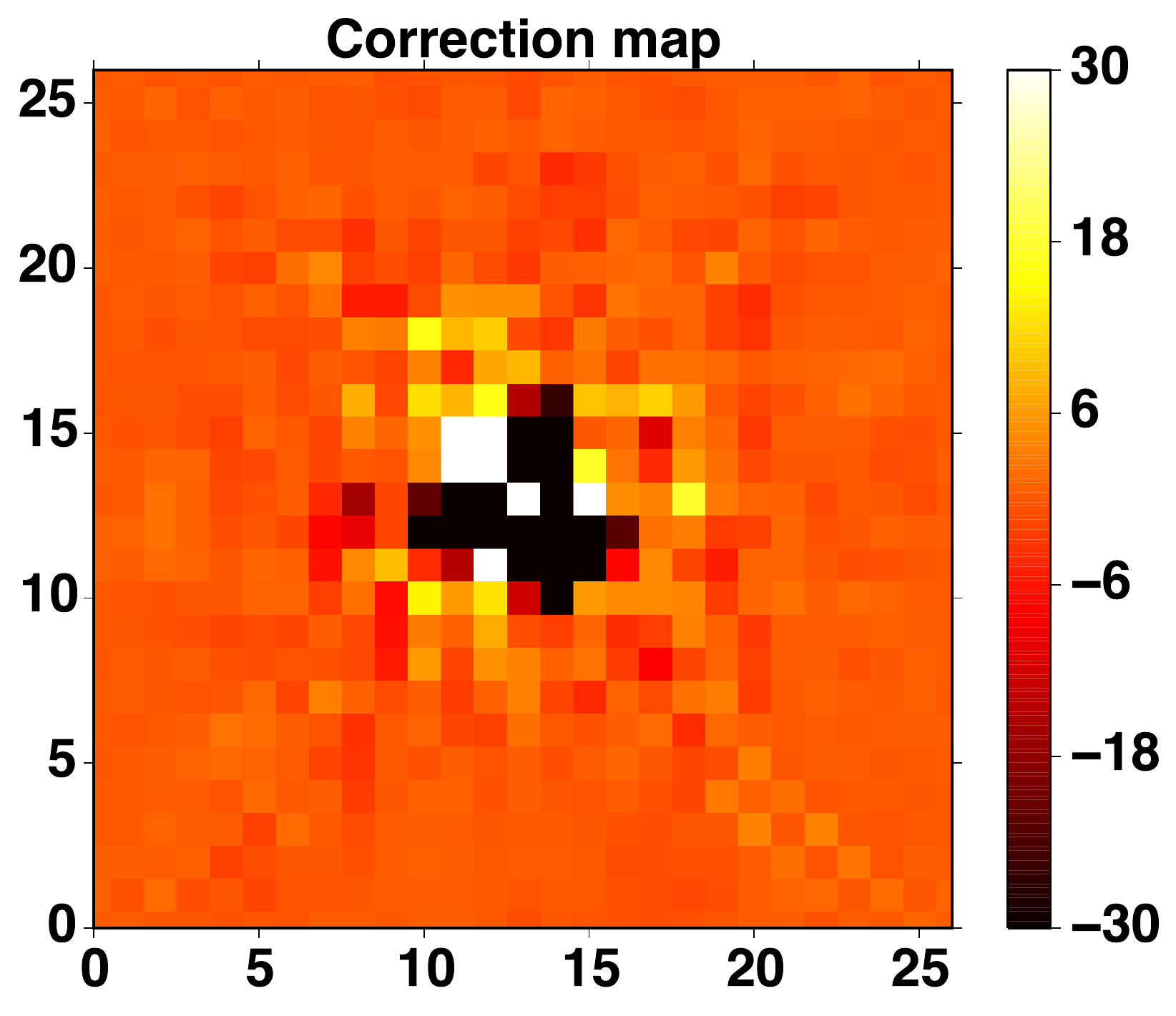}
  \caption{Point spread function subtraction allows isolating the
    secondary and accurately measuring its brightness in our WFC3
    F160W images. {\em Upper row:} original (left) and primary
    subtracted (right) images. After subtraction of the primary PSF
    and correction map, 2M1207b is detected at a high significance
    level. \revise{{\em Lower row:} examples of \tinytim{} PSFs for
      the primary and the secondary, and the correction map.}
  }
  \label{fig:1}
\end{figure*}

Presence of condensate clouds is one the most unique features of the
ultra-cool atmosphere of directly imaged exoplanets and brown
dwarfs. Studies of formation and properties of condensate clouds
\citep[e.g.][]{Ackerman2001, Burrows2006a, Helling2008, Allard2012}
have made great progress on the cloud behaviors across different
spectral types, especially the role that clouds play in the L-T
transition \citep[e.g.][]{Burrows2006a, Marley2010}.  Surface gravity
is suggested to be the second key parameters in defining cloud
structures \citep[e.g.][]{Marley2012} after effective temperature. Low
surface gravity objects (e.g. HR8799 bcd, \cite{Marois2008a}, 2M1207b, \cite{Chauvin2004}) are significantly redder and under-luminous
compared to field brown dwarfs.  The anomalous color and luminosity of
low surface gravity objects support models including unusually thick
clouds \citep{Currie2011, Madhusudhan2011,Skemer2011,
  Skemer2012}. However, due to lack of observational constraint, the
dependence of cloud properties on surface gravity is not very well
modeled.

Intensity modulations introduced by heterogeneous clouds can be
directly observed and studied via time resolved observation and
rotational mapping
\citep[e.g.][]{Apai2013,Buenzli2012,Buenzli2015,Burgasser2013,Radigan2012,Yang2015,Metchev2015,Heinze2015,
Biller2015}. These techniques isolate the effect of cloud
properties and obtained great success in determining the rotation
period and unveiling the structures of the atmospheres of brown dwarfs.
 \citet{Kostov2013} demonstrated that these techniques can be applied to directly imaged
exoplanets, too, allowing comparative studies of objects with
different surface gravities. However, high contrast amplifies the
challenges for directly imaged exoplanets and planetary-mass
companions to acquire high-precision light curves compared to
\revise{isolated} brown
dwarfs.

2M1207b \citet{Chauvin2004} is the first directly imaged extra-solar
planetary-mass companion. \citet{Chauvin2005} and \citet{Song2006}
confirmed that 2M1207b and its host 2M1207A form a bound, co-moving
system. 2M1207A and b have an angular separation of $0.78''$, which
corresponds to a projected separation of 41.2 AU at a distance of 52.4
pc \citep[e.g.][]{Ducourant2008}. Combining 2M1207b's age and near
infrared luminosity with brown dwarf cooling models
\citep[e.g.][]{Baraffe2003}, the object's mass is estimated to be
2.3-4.8 M$_{\mathrm{Jup}}$ \citep{Barman2011b}. Even though a
circumsubstellar disk was discovered around 2M1207A
\citep[][]{Sterzik2004}, the high companion-to-host mass ratio and
large separation argue for binary-like gravitational fragmentation
formation \revise{\citep{Lodato2005,Mohanty2007}}.

Early observations revealed that 2M1207b's color is much redder and
its near-infrared luminosity is much lower than those of field brown
dwarfs with similar spectra\citep[e.g][]{Mohanty2007, Skemer2011,
  Barman2011b}. 2M1207b's luminosity -- as derived from near-infrared
photometry -- is $\sim2.5$ mag lower than that predicted based on its
mid- to late L spectral type and effective temperature of $\sim 1600$
 K\citep{Patience2010}.  Based on multi-band, near-infrared photometry,
\citet{Skemer2011} argued that the apparent under-luminosity of 2M1207
b could be explained by a model of a spatially heterogeneous atmosphere
composed of patches of thin and patches of unusually thick clouds.
Similarly, \cite{Barman2011b} argued that non-local chemical
equilibrium could play an equally important role as thick clouds in
defining 2M1207b's color and luminosity.

The discovery of additional planetary-mass companions with similarly
red colors \revise{(e.g. AB Pic B, \citealt[][]{Chauvin2005b}, HR8799bcde,
  \citealt[][]{Marois2008a,Marois2010})} and apparent under-luminosity have
highlighted 2M1207b as a template of low gravity ultra-cool
atmospheres but as of now understanding the composition and structure
of clouds and their gravity-dependence remained elusive.

In this {\em Paper} we present the first, high-contrast, high-cadence,
high-precision, time-resolved {\em Hubble Space Telescope} (HST)
photometric time series of 2M1207b, a directly imaged planetary-mass
object. We successfully detect rotational modulation and measure the
amplitudes in two bands and determine the rotational period. These
observations probe the spatial heterogeneity and vertical structure of
clouds in planetary mass objects for the first time.

\section{Observation}

We obtained direct images of the 2M1207A+b system on UT 2014 April 11
from 08:07:47 \revise{(JD 2456758.838738)} to 16:53:18 \revise{(JD
  2456759.203681)} using HST and its Wide Field Camera 3 \citep[WFC3,
pixel scale=$0.130$mas/pixel,
][]{Mackenty2008} in the frame of the HST Program GO-13418 (PI:
D. Apai). We acquired the observations in filters F125W
($\lambda_{\mathrm{pivot}}$
= 1245.9 nm, full width at half maximum (FWHM) = 301.5 nm) and F160W
($\lambda_{\mathrm{pivot}}$
1540.52, FWHM = 287.9 nm), roughly corresponding to the J and H
bands. We used the $256\times256$
pixels sub-array mode to avoid memory dumps during the observations.
In order to provide a near-continuous coverage for detecting
modulations we observed the 2M1207 system in 6 consecutive HST orbits,
obtaining data with \revise{maximum} cadence of 1.78 minutes over a
baseline of 8 hours and 40 minutes. The observations were interrupted
by 58-minute long Earth occultations every 94 minutes.

The observations applied space craft rolls each two orbits to allow
roll-subtraction of the primary \citep[e.g.][]{Song2006}. The
telescope roll angles for orbit 1, 3, and 5, and those for 2, 4, and 6
differed by $25^{\circ}$. At the separation of 2M1207b, this angle
difference corresponds to a displacement of $0.34''$, or 2.75 and 2.30
resolution elements in F125W and F160W, respectively.
%In each orbit we took 8 SPARS10 exposure sequence with NSAMP=10,
%alternating between F160W and F125W filters, with 2--3 identical
%exposures in each exposure sequence.

\reviseTwo{In each orbit, we used the visibility of 2380s, which allowed us
  to take eight SPARS10 exposure sequences
  alternating between the F125W and F160W filter. Each sequence contained
  2--3 identical exposures of 88.4 s with 10 non-destructive read-outs. The number of exposures were limited by
  time spent on filter switching and transferring the data from the
  instrument to the data processing computer.} To improve sampling and reduce the risk
that the core of point spread function (PSF) is affected by bad
pixels, we applied a 4-point dither pattern with differential ``X/Y''
offsets of 1.375" in the detector frame, providing optimal
non-integral (half pixel) step of 10.5 and 8.5 pixels in F125W and
F160W, respectively. In total, we obtained 70 and 64 images  in
F125W and F160W, respectively.

\section{Data Reduction}

 \begin{figure*}
  \centering
  \plotone{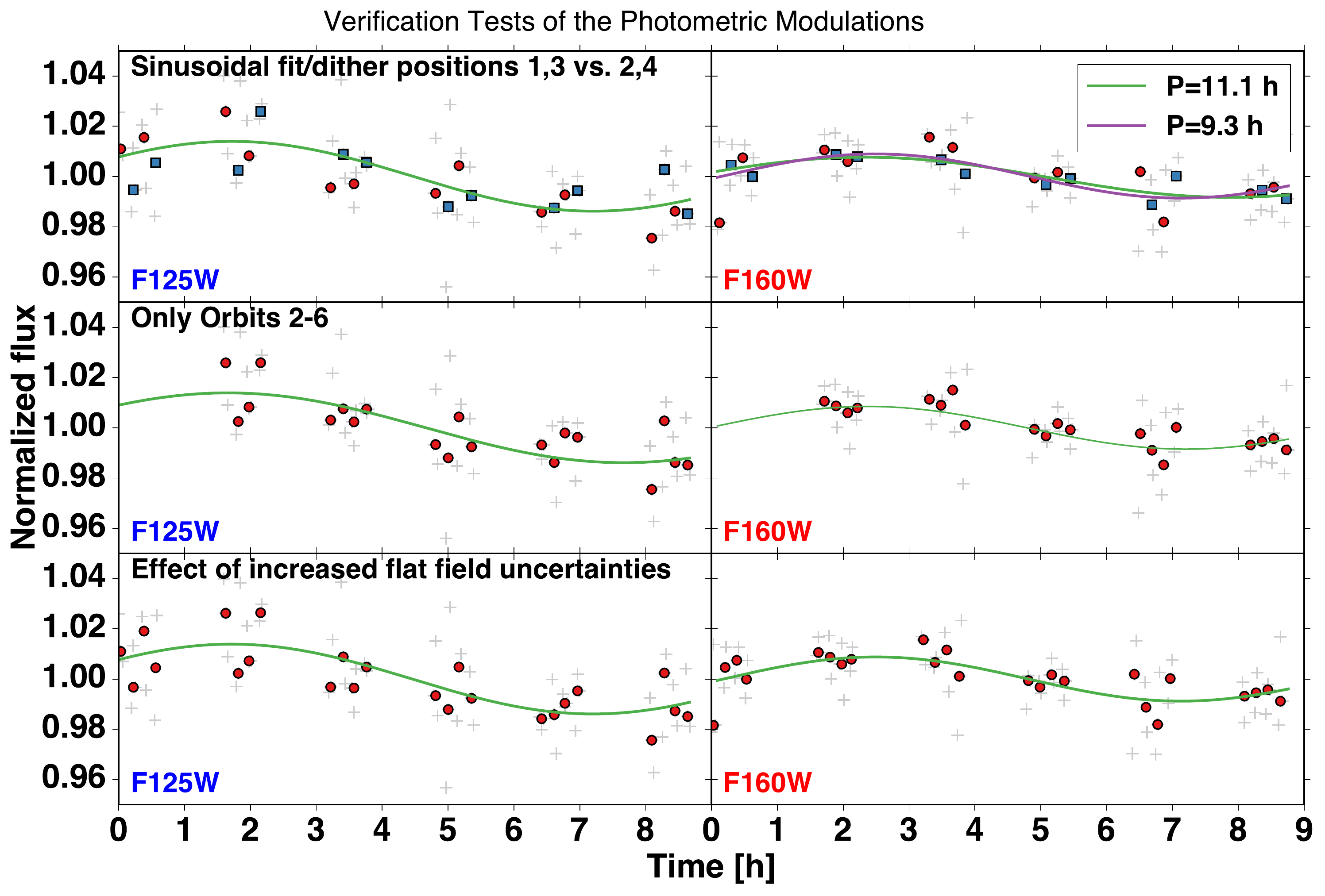}
  \caption{F125W (left) and F160W (right) light curves under different
    variability verification tests. Individual measurements are
    plotted with gray crosses. Photometric measurements of the same
    exposure sequence are binned, and binned photometry are plotted
    with points or squares. Best fitted sinusoids are plotted with
    solid lines. {\em Upper}: binned measurements taken in dithering
    position 1 and 3 (red points) and that taken in 2 and 4 (blue
    squares) are plotted with different symbols. \reviseTwo{The two halves
      of data were reduced independently, and show the same modulation
      trend.} In upper left panel, the green line a sinusoid fitted
    with all parameters freely varying, and the purple line is a
    sinusoid fitted with the period set the same as that of
    F125W. {\em Middle}: sinusoids fitted without using the data taken
    in Orbit \#1. These curves are almost identical to the curves
    plotted in upper panel. {\em Lower}: photometry measured with
    AFEM-added images and best fitted sinul curves. These points and
    curves are also almost identical to those plotted in the upper
    panel. \reviseTwo{The slight difference of normalized photometry
      among panels for the same filters is the result that
      normalization factors change when omitting the first orbit or when
      adding artificial noise.}}
  \label{fig:2}
  
\end{figure*}

\subsection{Photometry}

We started the reduction from the \flt{} produced by the WFC3's
\texttt{calwfc3} pipeline. We did not opt to use \ima{} that contain
all non-destructive read-outs, because they provided less information
on 2M1207A, which saturated after the first few samples.  The \flt{}
are results of basic calibration, including dark current correction,
non-linearity correction, flat field correction, as well as
up-the-ramp fit on the non-destructive read-outs. 
Pixels with data quality flags ``bad detector pixels'', ``unstable
response'', and ``bad or uncertain flat value'' were masked out and
excluded from further analysis as suggested by previous transit
exoplanet spectroscopic observations \citep[e.g.][]{Berta2012,
  Kreidberg2014}.

The major challenge of high contrast observation with WFC3/IR is the
fact that the detector is significantly under-sampled.  2M1207A and b
are only separated by $\sim6$ pixels or $\sim$5 FWHM of the PSF on the
detector. When applying roll subtraction, notable artifacts are
introduced by image shifting and interpolation.  \tinytim{} PSF
simulator \citep{Krist1995} offers a solution by providing Nyquist or
better sampled PSF, but systematic errors of \tinytim{} PSF for WFC3
limits its ability in high precision photometry \citep{Biretta2014}.
Building on the large number of PSFs obtained in our program at two
different roll angles, we followed a novel, two-step approach that
uses a hybrid PSF.  First, based on our observations, we derived
correction maps for \tinytim{} that accurately described the scattered
light component for the primary at the correct location on the
detector. Second, we carried out a PSF-photometry using hybrid PSFs
composed by \tinytim{} PSFs and the correction map by simultaneously
minimizing the residuals from the primary and the secondary.

For both of 2 steps, we used \tinytim{} to calculate 10$\times$
over-sampled model PSFs based on the filters, the spectra (2M1207A:
\cite{Bonnefoy2014}, 2M1207b: \cite{Patience2010}), the telescope's
actual focus, and the telescope jitter.  We used the set of \tinytim{}
parameters provided by \cite{Biretta2014} to improve modeling the cold
mask, diffraction spikes, and the coma. The focus parameters are 
interpolated to the precise time of the observations using the
tabulated values provided by
STScI\footnote{\url{http://www.stsci.edu/hst/observatory/focus/FocusModel}}.
To align the \tinytim{} PSF to the observed PSF of 2M1207A, we
moved the over-sampled PSF on a coordinate grid (gird size=0.001 pixel)
using cubic interpolation, and searched for the position that minimizes
the rms difference of the observed and the re-binned \tinytim{} PSF over
a region centered on 2M1207A with a 5-pixel-radius aperture centered
on 2M1207b excluded.  Then we introduced another \tinytim{} PSF for
2M1207b and fit the position of 2M1207b and the scales of the
\tinytim{} PSFs of 2M1207A and b simultaneously by minimizing the
residual from both primary and secondary. In the first step, we discovered that the
difference of observed PSFs and model PSFs was very stable for a
specified telescope roll angle and dithering position. Therefore, at the end of the first step,
we derive 8 (2 roll angles $\times$ 4 dithering positions) correction
maps for each filter:
\begin{equation}
  \mathrm{Corr = Median(PSF_{obs.} - PSF_{model} )}
\end{equation}
where $\mathrm{PSF_{model}}$ was a combination of two scaled \tinytim{} PSFs
for 2M1207A and b. In the second round, we combined the correction
map linearly with the two \tinytim{} PSFs to generate hybrid PSFs,
and scaled the correction map together with the two PSFs so that the
residual, which is expressed as
\begin{equation}
  \mathrm{Residual} = \mathrm{Image} - a\times \mathrm{PSF_{A}} - b\times
    \mathrm{PSF_{b}}- c\times\mathrm{ Corr}
  \end{equation}
  is minimized by least square fitting.
We found that by introducing the correction term,
the reduced $\chi^{2}$ decreased from $\sim 10$ to
$\sim 1$. Relative photometry was acquired from the scaling
parameters of the \tinytim{} PSFs.

Our final step was to correct for a slight apparent trend between the
position of the targets on the detector and their fluxes. We attributed
this to a combination of slight changes in the PSF profile due to
pixelation and to the effect of imperfectly corrected pixel-to-pixel
sensitivity variations. We corrected for the apparent
position-dependent flux changes by normalizing each photometric point
by the median of all fluxes measured when the target was at the
same position, i.e. combining data over 6 orbits.  We note that this
correction was small and, as we demonstrate in the next sections,
could not introduce artificial modulations that resemble the long-period
variations that we identified in 2M1207b.

As our study is the first to present high-contrast, high-cadence
observations, we provide a detailed analysis of the uncertainties and
their impact on our results.

\subsection{Uncertainty Analysis: White noise}

First we estimated the photon noise for the photometry of
2M1207b. The total photon noise of the photometry was calculated by
combining the photon noise of every pixel, which was derived
from count rates and detector gain. The photon noises 
in F125W and F160W are 1.33\% and 1.02\%, respectively.

Since the PSFs for the 2M1207A and b were fitted simultaneously, the
uncertainties for photometry and position of the primary and secondary
were coupled. Errors in position measurements of 2M1207A could
potentially affect the photometry of 2M1207b. We used a Monte Carlo
(MC) method to evaluate the overall systematics of the PSF fitting. We
applied photometry to images that were added with random
Poisson noise and repeated the photometry procedure for 1000 times. The
uncertainties for F125W and F160W photometry were found to be 1.34\% and 1.12\%,
respectively. 

\subsection{Uncertainty Analysis: Flat field uncertainties}

A further contribution to photometric uncertainties may be introduced
by imperfectly corrected pixel-to-pixel sensitivity
differences. 2M1207b were observed at 8 different positions on the
detector (2 rolls $\times$ 4 dithering positions). Imperfect flat
field correction could introduce position-dependent
differences in the count rates. The uncertainty of WFC3 IR flat field
is typically $\sim 1\%$ \citep{dressel2012wide}.

In PSF photometry, however, multiple pixels are fitted simultaneously,
so that we expect the photometry to be less affected by high spatial frequency flat field
noise, and have a lower than 1\% uncertainty from the flat field
errors. To verify this, we multiplied every image by an artificial flat
field error mask (AFEM) -- a uniformly distributed Gaussian noise array with
mean of 1 and sigma of 1\% -- and repeated the PSF photometry on the
resulting images.  The analysis of these experiments resulted in almost
identical light curve to the original, verifying that the flat field
errors did not affect our photometry significantly (Figure
\ref{fig:2}, bottom panel).

  \begin{figure*}
  \centering
  \plotone{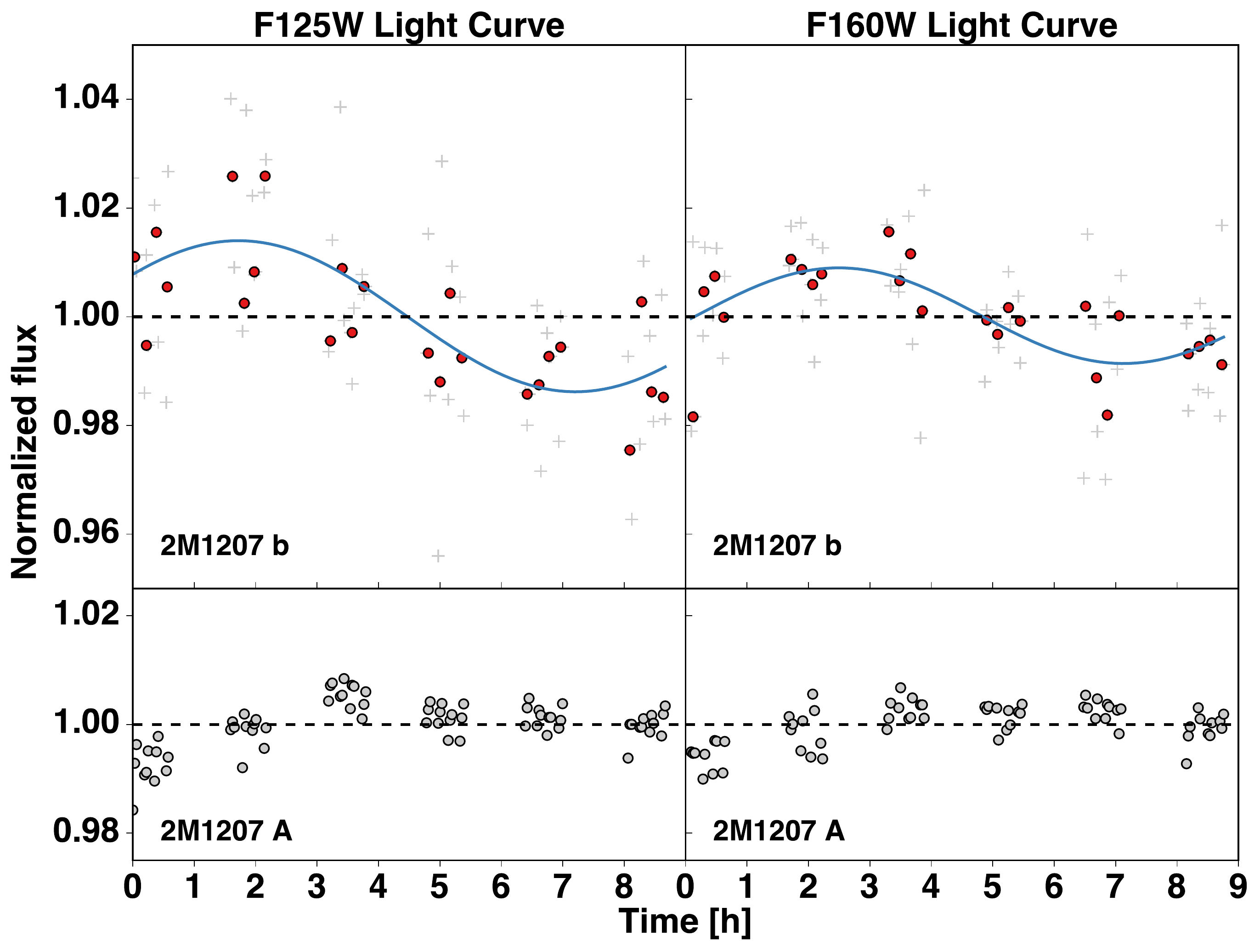}
  \caption{Normalized light curves for 2M1207 B (upper) and A (lower)
    with filter F125W (left) and F160W (right). Unbinned
  measurements are plotted in gray crosses and binned photometry are
  plotted with red points. Best fitted sinusoids are plotted
  with blue solid lines.}
  \label{fig:3}
\end{figure*}

\section{Verification of Photometric Modulations and Amplitude
  Estimate}

\subsection{Tests and Verification}

The light curves that resulted from our photometry showed apparently
sinusoidal modulations, discussed in more detail in
\S\ref{Results}. To verify that these modulations are intrinsic to
the object and not the result of our data reduction procedures or 
instrumental changes, we carried out three different tests.

First, we fitted sinusoids independently to the light curves of two
filters to verify the similarity of the signal in the two bands
(Figure \ref{fig:2}, top panel) \reviseTwo{using an Markov Chain Monte Carlo (MCMC)
 approach (for detail see \S\ref{sec:MCMC})}. Inconsistent periods or light curve
shapes would argue against a genuine signal. \reviseTwo{We used
  sinusoids as examples of the simplest periodic functions.} We found that the periods
of the best fit sinusoids were similar, \Jperiod{} h for
F125W and \Hperiod{} h for F160W. These periods are consistent
within the uncertainty. Furthermore, these periods are not close to
any timescales over which HST or WFC3 changes, and are
very different from all timescales present in our observations
(dithering timescales, integration times, and orbital
timescales). \reviseTwo{We note that the period is close to
  the total time span of the observation, however, the probability that
  the period  is equal to or less than the observation time baseline is
  negligible (see \S\ref{sec:MCMC}).}

As a second test, we repeated the analysis neglecting the first
orbit. The motivation behind this test is that, due to spacecraft
thermal settling, the first orbits of HST observations are often
slightly unstable, and are neglected in high-precision studies
\citep[e.g.][]{Mandell2013}. Indeed, in our analysis, 2M1207A is
significantly fainter in the first orbit (Figure \ref{fig:3}) than in
the subsequent ones.  \reviseTwo{Our analysis based on orbits 2--6 found the
periods were \Jperiodd{} hours for F125W and \Hperiodd{} hours for F160W, }
which were almost identical results to our analysis using the whole 6 orbits, based on
which we conclude that the first less reliable orbit does not affect
our results significantly (Figure \ref{fig:2}, middle panel).

As a third test, we explored whether a subset of images, perhaps due to
imperfect normalization or correlation with specific instrument states,
could drive the light curves into apparently sinusoidal shapes. To
test this possibility, we split the data into two temporally
overlapping halves: subset one were images taken at dithering position
1 and 3; subset two were those taken at dithering position 2 and
4. For both subsets, we repeated our analysis independently.  For both
of F125W and F160W, two halves demonstrated similar sinusoidal
modulations.  Our analysis detect sinusoidal modulations in {\em both}
subsets and in {\em both} filters, with periods and amplitudes
consistent with those derived from the complete data set (Figure
\ref{fig:2}, upper panel).
 
 These tests demonstrate that the modulation seen in our data are
 consistently present in the different filters, in the different time
 segments of the data, and in data obtained in different dithering
 positions. All of the three tests support the signal to be
 intrinsic to the target.

\subsection{Amplitude and Period Measurements}
\label{sec:MCMC}

\begin{figure*}
  \centering
\plotone{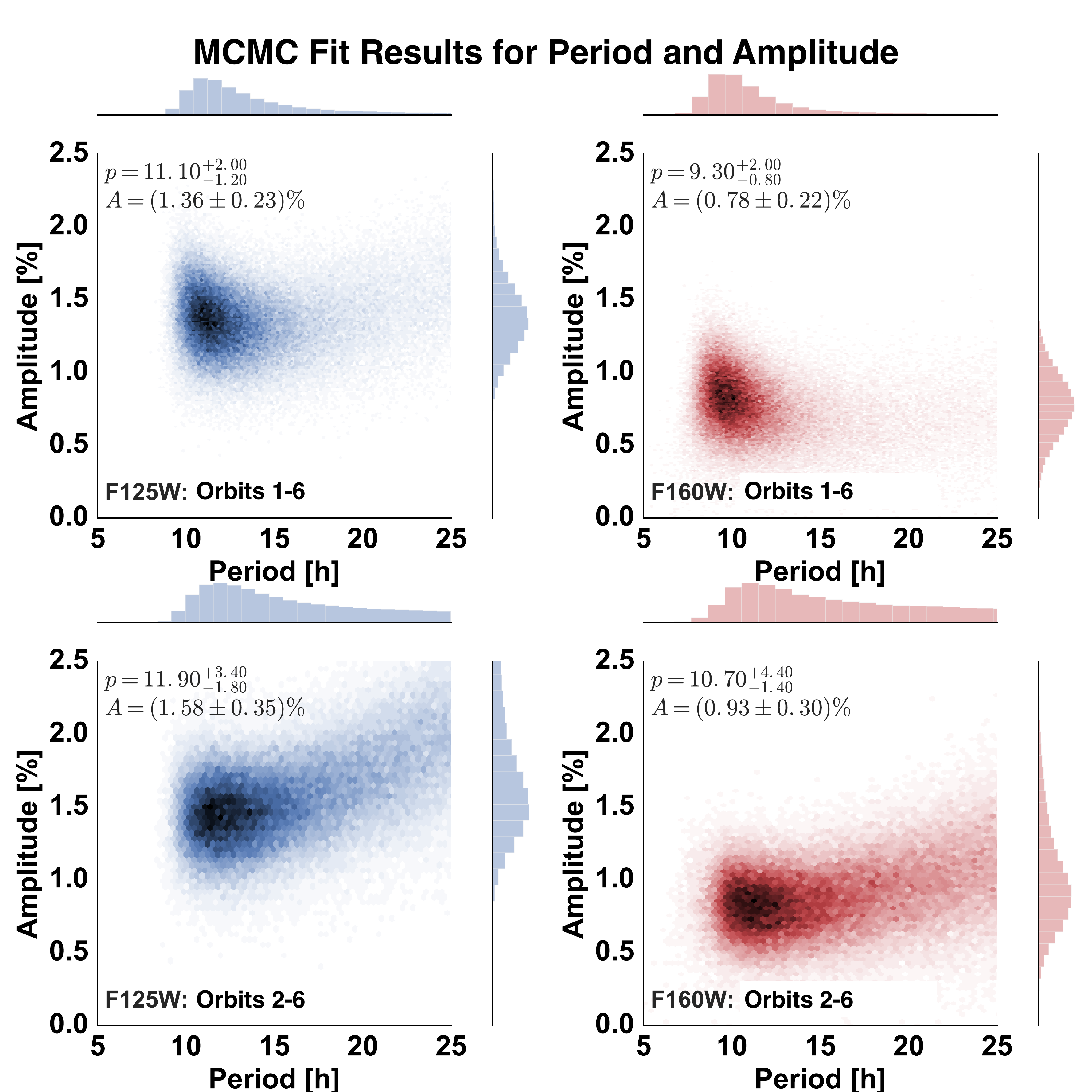}
  \caption{\revise{Posterior distributions for amplitudes and periods for F125W
    (left) and F160W (right) light curves, using data from orbits 1-6
    (upper) and data from orbits 2-6 (lower). In each panel,
    the univariate distributions for period and amplitude are plotted
    along side with the joint distribution. The values and
    uncertainties of period and amplitude are shown on the upper left
    corner of each panel.}}
  \label{fig:4}
\end{figure*}

\revise{To constrain the amplitudes and periods of the light curves,
  we performed sinusoidal fit using MCMC method.  The priors for the
  amplitudes, the periods and the phases of the sine waves were all
  assumed to be uniform distributions. The ranges of the prior
  distributions were 0 to 5\% for the amplitudes, 2 to 40 hours for
  the periods, and 0 to $2\pi$
  for the phases. The results of the MCMC fit are shown in Figure
  \ref{fig:4}. The posterior distributions of the periods and
  amplitudes are very well constrained for both two filters so that we
  are able to measure the amplitudes and periods as well as their
  uncertainties. For F125W, the amplitude and period are \Jamp{} and
  \Jperiod{} \reviseTwo{hours}, respectively, and for F160W, they are
  \Hamp{} and \Hperiod{} \reviseTwo{hours},
  respectively. \reviseTwo{Next, to place the strongest constraint on
    the period of 2M1207b, we jointly fitted the two light curves
    requiring an identical period for two filters but allowing
    different phases and amplitudes. The period measured from the
    posterior distribution of the joint fit is \period{} hours.} We
  also repeated the fitting excluding data from the first orbit in
  order to quantitatively assess the influence of these less stable
  data points. Even in the case of limiting our data to Orbits 2-6,
  the posterior probability distribution peaks at periods and
  amplitudes very similar to those found for the entire dataset,
  demonstrating that a periodic solution is preferred with or without
  the first orbit data. Nevertheless, unsurprisingly, when shortening
  the baseline by considering only Orbits 2-6 results in a long-period
  tail in the probability distribution.}

\reviseTwo{To evaluate the possibility of very long period sinusoids,
  we integrated the posterior distributions for periods at least
  2$\sigma$
  longer than the best fit period, as well as for periods longer than
  20 hours. For the joint fit, the integrated probabilities are
  12.68\% longer than and 1.29\% for the period longer than 13.1
  ($\mathrm{best\, fit + 2}\sigma$)
  hours and 20 hours, respectively. Given these results we conclude
  that periods much longer than our baseline are very unlikely.}

\section{Result}
\label{Results}

We present the first high-contrast, high-cadence, and high-precision
photometry of a directly imaged planet or planetary-mass
companion around another star. Our observations reveal a modulation in the light curve of
the $\sim 4 \mathrm{M_{{Jup}}}$ companion 2M1207b, the first detection
of modulations in directly imaged planetary-mass objects. The best
fit periods for F125W and F160W are $11.1_{-1.2}^{+2.0}$  and $9.3_{-0.8}^{+2.0}$ hours,
respectively. Jointly fitting the two light curves gives a period
measurement of \period{} hours.

We obtained high signal to noise photometry for both 2M1207A
and B (Figure \ref{fig:3}). On average, the photometric contrast is
$6.52\pm0.01$ mag for F125W and $5.77\pm0.01$ mag for F160W. We
provide our photometry result data in Table \ref{tab:F125W}, \ref{tab:F160W}.

We find that the amplitudes in the two bands are
significantly different. By fitting Gaussians to the MCMC fit result
distributions, we determin that the  amplitude for F125W is
1.36\% with a standard deviation of 0.23\%, and  the amplitude for F160W is
0.78\% with a standard deviation of 0.22\%. The amplitudes of two
bands are separated by more than 2-$\sigma$. The amplitude
for F125W is $1.74\pm0.30$ times of that for F160W light curve.

\section{Discussion}
\label{sec:discussion}
\begin{figure*}
  \centering
  \plottwo{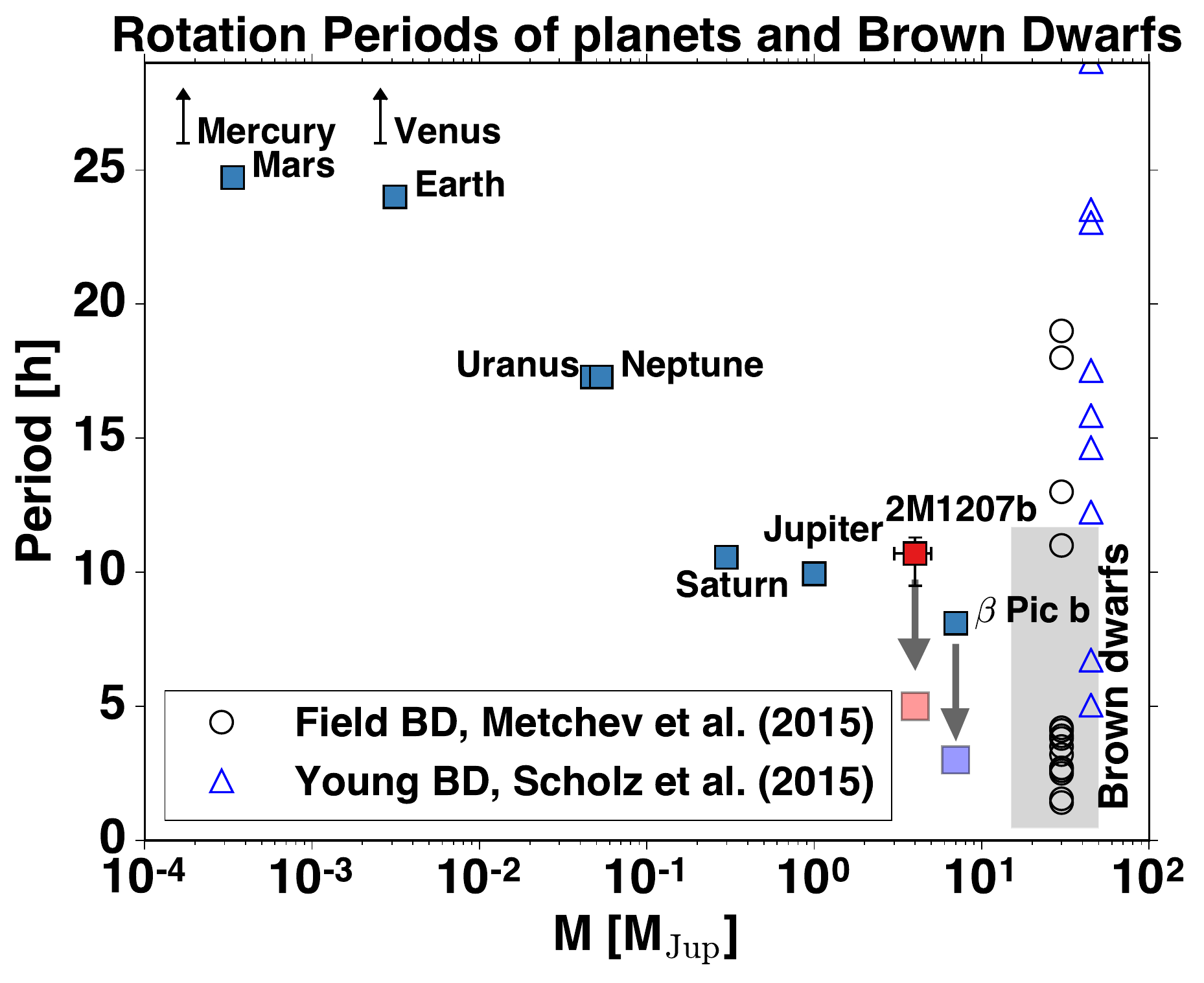}{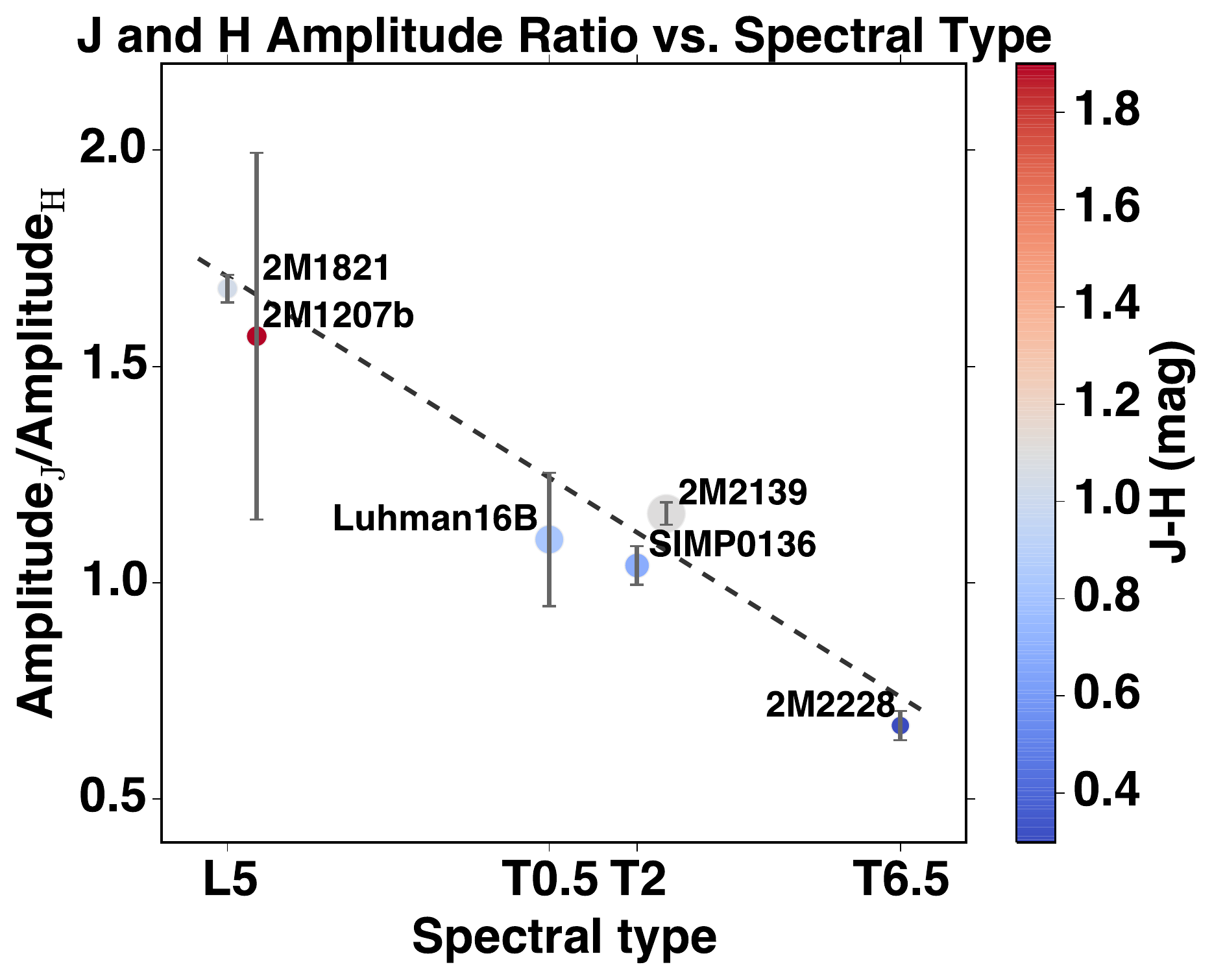}
  \caption{comparison of 2M1207's rotation period and color change
    with brown dwarfs, \bpic{} b, and solar system planets. {\em
      Left}: period vs. mass plot for 2M1207b (red square), solar
    system planets and \bpic{} b (blue squares), field brown dwarfs
    from the study of \citet[][]{Metchev2015}
    (black circles, gray shade) and young brown dwarfs from the study
    of \citet{Scholz2015} (blue
    triangles). Final rotation rates for 2M1207b and \bpic{} b
    estimated based on conservation of angular momentum are plotted
    with faint red and blue squares, respectively. The mass of brown
    dwarfs are assumed to be $\sim30$ \mjup{}, and the difference mass
    of field brown dwarfs and young brown dwarfs are added
    artificially for the sake of clarity . The gray rectangle that has a
    $\pm$15
    \mjup range in $x$,
    and a $\pm \sigma$
    of field brown dwarf periods range in $y$,
    indicates a region where field brown dwarfs most likely to appear in
    this diagram. Rotation period monotonically decreases with the
    increase of mass. {\em Right}: ratio of modulation amplitude in J
    and H band vs. spectral type for 2M1207b and brown dwarfs. The
    point for 2M1207b is shifted to +$x$
    for half spectral type for clarification.  The colors of the
    points represent J$-$H
    magnitude, and the sizes of the points are proportional to the
    J-band modulation amplitudes. The gray dashed line is the result
    of a linear fit to these points.  Correlation of J- and H-band
    modulation amplitude ratio and spectral type is shown.}
 \label{fig:5}
\end{figure*}

% A fundamental result of our study is the direct determination of the
% rotation period of a directly imaged planetary-mass object. We
% convert the rotation period to equatorial velocity  by adopting a radius of 1 -- 1.4
% $R_{\mathrm{Jup}}$ for 2M1207b, and 1 $R_{\mathrm{Jup}}$ for field
% brown dwarfs with well defined rotation period from the study of
% \cite{Metchev2015}, and compare their rotation velocities with solar
% system planets and \bpic{} b in the left
% panel of Figure \ref{fig:5}.  The study byf
% \citep[][]{Snellen2014} succeeded in measuring \vsini{} for \bpic{} b
% and demonstrated that it fits a trend defined by Solar System planets
% in which more massive planets have faster rotation rates. They suggested
% that this relation is linked to the accretion processes during planet
% formation.
\reviseTwo{The baseline of our observations is not long enough to
  cover a complete period of 2M1207b. This should come as no surprise,
  given the 10-20 hour long periods observed in the Solar System giant
  planets. Nevertheless, as the observations may cover a large
  fraction of the complete rotation of 2M1207b, we can derive the
  probability distribution of the rotation period and the amplitudes
  (in two bands) based on our data. To do this, we assumed an example
  of a simplest periodic functions, a sine wave. Approximating
  periodic or quasi-periodic modulations in light curves with moderate
  signal-to-noise with sine waves is a common approach \citep[e.g.][]{Buenzli2012}.  We
  used this approximation and combine it with an MCMC approach in
  \S\ref{sec:MCMC} to derive posterior probability distributions.}

\reviseTwo{Before discussing our MCMC results for completeness we
  discuss why other, non-periodic functions cannot be considered
  adequate fits to our data. We can exclude a flat (zero slope)
  lightcurve due to its poor fit (reduced $\chi^{2}$=117.3/69,
  76.9/63 for F125W and F160W, respectively). A constant, but non-zero
  slope line provides an improved reduced $\chi^{2}$,
  but is not a physically viable model: 2M1207b's time-averaged
  brightness cannot rapidly increase or decline. A linear decline or
  brightening that is not intrinsic but instrumental in nature can be
  excluded for three reasons: 1) the simultaneously observed 2M1207A
  shows no such modulation; 2) the signal is consistent between
  different filters and different
  parts of the dataset; 3) all known HST thermal responses occur on
  timescales of 1 orbit or shorter \citep{Lallo2005}, inconsistent
  with the modulations in our data. Given that the linear models are
  inconsistent with our data or not physically viable, and the sine
  waves used in our MCMC fits provide lower reduced $\chi^{2}$ than any of
  the linear models in both filters, we adopt those as the simplest
  description of the modulations.  We include sine waves of period as
  long as 40 hours as possible solutions for our MCMC fit to consider
  the possibility of very long period variations. We find that a
  sinusoid with a period of $\sim 11$
  hour is the most probable solution with or without including data
  from the first orbit. When jointly fitting F125W and F160W light
  curves with sinusoids with the same period but independent amplitudes and
  phases, the integrated posterior distributions for periods at least 2-$\sigma$
  larger than the best fit period is 12.7\%, and that for period
  longer than 20 hours is only 1.3\%. Therefore, our observations exclude
  the possibility of a very long period (p$>$20 hours) at a high
  confidence level. Nevertheless, the period measurement should be further
  constrained in the future  when light curves with full phase
  coverage are available.}

The direct measurement of the
photometric modulation period of a directly imaged planetary-mass
object is an import result of our study.
% \revise{Although a linear trend can also provide reasonable fit to
%   the light curves when data from the first orbit are excluded, it is
%   not a physical viable model. The light curve of a planetary mass
%   companion is either a complete flat line, or a periodical curve with constant mean. On the contrary,
%   2M1207b is not likely to constantly getting bright or faint. The
%   flatness of the light curves of 2M1207A demonstrate that there is no
%   instrumental sensitivity drift or any similar effect that can
%   introduce a $\sim9$ hour slope. We do include very 
%   long period sinusoid as possible solution for our MC fit, and
%   demonstrate that a sinusoid with period of $\sim10$
%   hour is the most probable solution with or without including data
%   from the first orbit.}
We infer the rotation period of
2M1207b to be the same as the period of photometric
variation. Although horizontal winds can cause the measured period to
differ from true rotation period, it is unlikely that the difference
is greater than the uncertainty of the measurement, since for Jupiter
and Saturn, typical wind speeds are more than one order of magnitude
smaller than the equatorial rotation speeds.
% Our
% inferred rotation period of 2M1207b is typical among those of brown dwarfs
% \citep{Reiners2008}.
In the left panel of Figure~\ref{fig:5} we compare the rotation period
of 2M1207b to the solar system planets, \bpic{} b the only other
directly imaged planet with an estimated period and measured \vsini,
field brown dwarfs from the study of \citet[][]{Metchev2015},
\revise{and young brown dwarfs (UpSco, age $\sim 11$ Myr) from
  \citet[][]{Scholz2015}}. \citet[][]{Snellen2014} measured \vsini{}
for \bpic{} b and demonstrated that it fits a trend defined by Solar
System planets in which more massive planets have faster rotation
rates. The interesting finding that \bpic{} b, an exoplanet that
formed in a protoplanetary disk, follows this trend suggests a
possibly connection between planet mass, initial angular momentum, and
formation in a disk.

Excitingly, our measurement of the rotation period of 2M1207b, a
planet mass companion that has similar age to \bpic{} b, has a
rotation period that fits in the same trend, as well as majority of
the brown dwarfs. As 2M1207b and \bpic{} b evolve and cool down, they are
expected to shrink to the size of Jupiter. Order of magnitude
estimation based on the conservation of angular momentum results in
final rotation periods of $\sim 5$ h and $\sim 3$ hours
\citep{Snellen2014} for 2M1207b and \bpic{} b, respectively, which
still fit to the period vs. mass trend. Although 2M1207A is known to
host a circumsubstellar disk \citep{Sterzik2004}, the low mass of
typical brown dwarf disks \citep[e.g.][]{Klein2003, Mohanty2013} and its
large separation argue against the possibility that 2M1207b has formed
in a protoplanetary disk.  The result that objects formed in different
scenarios share the same trend of period vs. mass suggests that
rotation periods -- in absence of well-determined ages -- may
contribute insufficient evidence for a formation in a disk vs. in a
cloud core environment.

\revise{ The rotation period of $\sim 10$
  hour is significantly longer than those of field brown dwarfs from the sample
  of \cite{Metchev2015}, and the corresponding equatorial rotation
  speed ($\sim 15\, \mathrm{km\,s^{-1}}$) is lower than most L-type field
  brown dwarfs \citep[][]{Reiners2008}.
  In contrast, the rotation period of 2M1207b is similar to the median period
  of the sample of \cite{Scholz2015}, whose age is similar
  to 2M1207b. The rotation period of 2M1207b
  is within the range predicted by evolutionary track
  established from measurements of brown dwarfs assuming conservation
  of angular momentum \citep[see Fig. 4 of ][]{Scholz2015}, and much longer
  than the break-up limit \citep{Marley2011}, i.e. the rotation period where the
  equatorial centrifugal force exceeds gravitational force. \reviseTwo{Observed
  rotation rates for brown dwarfs of show little
  evidence for angular momentum loss 
  \citep{Bouvier2014, Scholz2015} for the first few Myr of evolution, and agree with the model of solid
  body rotation. In contrast, young low
  mass star experience strong angular momentum loss and internal
  angular momentum redistribution at similar ages.} Rotation periods of
planetary mass objects with well established age measurements can
\reviseTwo{place  further constraint on the gravitational contraction and
  angular momentum evolution in the planetary mass regime.}}

Our inferred rotation period is very similar to those of Jupiter and
Saturn, which have periods of 9.9 and 10.5 hours, respectively.
Moreover, our inferred rotation is sufficiently fast that -- just as
with Jupiter and Saturn -- the atmospheric dynamics is likely to be
rotationally dominated at regional to global scales
\citep[][]{Showman2013}.  The importance of rotation can be
characterized by the Rossby number $Ro = U/\Omega L$, where $U$ is the
characteristic wind speed, $\Omega$ is the angular rotation rate
($1.7\times10^{-4}\rm\,s^{-1}$ for a 10-hour rotation period), and $L$
is the characteristic horizontal length scale.
\citet[][]{Showman2013} presented a theory of the
atmospheric circulation on brown dwarfs and directly imaged giant
planets, which predicts wind speeds in the range of tens to hundreds
of $\rm m\,s^{-1}$ depending on parameters. Using similar arguments,
\citet[][]{Apai2013}  argued for wind speeds of a few hundreds $\mathrm{m\,s^{-1}}$
(somewhat faster than typical in Jupiter) in two L/T transition brown
dwarfs.
Global simulations of the
atmospheric circulation using a one-layer model by \citet{Zhang2014}
predict a similar range.  Considering wind speeds ranging from 10 to
$1000\rm\,m\,s^{-1}$, a circulation that is global in scale
(L=$R_{\mathrm{Jup}}$) implies Rossby numbers of 0.001 to 0.1 on
2M1207b (see Figure 1 in \citet{Showman2013}).  For a circulation
whose length scale is 0.1 $R_{\mathrm{Jup}}$, the Rossby numbers would
range from 0.01 to 1 depending on wind speed.  Thus, over almost the
full range of plausible parameters, we expect that the large-scale
circulation on 2M1207b --- like Jupiter, Saturn, and most brown
dwarfs --- exhibits a Rossby number much less than one.  This implies
that the atmospheric circulation is rotationally dominated and that
the horizontal force balance is approximately geostrophic, that is, a
balance between Coriolis and pressure-gradient forces.

From the perspective of atmospheric dynamics, 2M1207b exhibits other
important similarities to brown dwarfs.  Its high effective
temperature indicates that -- like most brown dwarfs -- 2M1207b exhibit a
strong interior heat flux presumably transported by convection, and
that, by comparison, the external irradiation is negligible to the
circulation.  These similarities suggest that the overall dynamical
mechanisms for driving an atmospheric circulation on 2M1207b should be
similar to those on brown dwarfs.  Thus, given the prevalence of
infrared light curve variability observed on brown dwarfs, it is expected
to find such variability on directly imaged planets like 2M1207b.
Nevertheless, directly imaged planets generally have lower surface gravity
than field brown dwarfs, and this will affect the details of the
atmospheric circulation, potentially including the cloud patchiness.
Further observations of 2M1207b and other directly imaged planets will
help to elucidate these differences.  Key questions for the future
will include assessing the extent to which the atmospheric circulation
on 2M1207b -- including the existence of absence of zonal (east-west)
jet streams, vortices, storms, and turbulence, and their effect on
cloud patchiness -- are similar or different than that on typical field
brown dwarfs.

Our observations also allow us to compare the relative amplitudes in
the J- and H-bands with the handful of brown dwarfs for which
high-quality near-infrared time-resolved observations have been
obtained. In the right panel of Figure \ref{fig:5}, we compare the
relative amplitude of J- and H-bands of 2M1207b and brown dwarfs
\citep{Apai2013,Buenzli2012,Buenzli2015,Yang2015} that have different
spectral types and J$-$H colors.  J and H band fluxes for brown dwarfs
are integrated from WFC3 grism spectra using standard J and H filter
transmission profile. We find an interesting possible correlation
between the spectral types of the objects and their J- to H-band
amplitude ratios. In the right panel of Figure \ref{fig:5}, we show
that earlier spectral type objects have larger amplitudes at shorter
wavelength than at longer wavelengths. Interestingly, although the
J$-$H color of 2M1207b is significantly redder, its relative amplitude
ratio is very similar to that of 2M1821, which also has an L5 spectral
type \citep{Yang2015}.  This exciting, but tentative trend must be
confirmed with a larger sample of sources that also sample a broader
range of surface gravities as well as spectral types.  If the larger
sample verifies the trend suggested by our small sample, the amplitude
ratio will provide a powerful probe of the spectral type and surface
gravity dependence of vertical cloud structure.

\revise{Recently, \citet{Karalidi2015} showed that an MCMC-optimized
  light curve modeling tool can correctly retrieve two-dimensional
  atmospheric features from high quality light curves. In the future,
  with higher signal-to-noise light curves of planetary mass objects
  and exoplanets we will be able to map their atmospheres in greater
  detail.}

\section{Conclusions}
In summary, from our J- and H-band high precision, high-cadence light
curves we discovered sinusoidal modulations in the planetary-mass
object 2M1207b. This is the first detection of rotational modulations
in a directly imaged planetary-mass object.  \reviseTwo{By fitting the
  sinusoids to the light curves, we find a period of \period{} hours
  that is 20\% longer than our observation baseline and should be
  further constrained with full phase coverage in the future.} The
10.7-hour period is similar to that derived from \vsini{} 
measurements for the directly imaged exoplanet \bpic{} b.
\revise{The period of 2M1207b is longer than most field brown dwarfs
  with known rotation period,  but is similar to brown dwarfs
in a sample with an age similar to that of 2M1207b.} The amplitude ratio of J- and H-band is very
similar to that of a field brown dwarf with identical L5 spectral type, although they have
very different J$-$H colors.

Finally, we note that the observations presented here open an exciting
new window on directly imaged exoplanets and planetary-mass
companions. Our study demonstrates a successful application of
high contrast, high-cadence, high-precision photometry with planetary
mass companion. We also show that these observations can be carried
out simultaneously at multiple wavelengths, allowing us to probe
multiple pressure levels. With observation of a larger sample and at
multiple wavelengths, we will be able to explore the detailed
structures of atmospheres of directly imaged exoplanets, and identify
the key parameters that determine these.

\acknowledgments

We thank the anonymous referee for the suggestions that are helpful
for improving the manuscript. Support for program number 13418 was
provided by NASA through a grant from the Space Telescope Science
Institute, which is operated by the Association of Universities for
Research in Astronomy, Inc., under NASA contract NAS5-26555. The
results reported herein benefited from collaborations and/or
information exchange within NASA's Nexus for Exoplanet System Science
(NExSS) research coordination network sponsored by NASA's Science
Mission Directorate. M.S.M. acknowledges support from the NASA
Astrophysics Theory Program. A.P.S. acknowledges support from NSF
grant AST1313444.

% \bibliography{ref.bib}

\clearpage
\appendix
\LongTables
\begin{deluxetable}{cccccccc}
\tabletypesize{\scriptsize}
\tablecaption{F125W Photometry Results\label{tab:F125W}}
\tablewidth{0pt}
\tablehead{
\colhead{Orbit} & \colhead{Pos. Angle} & \colhead{Dither} & \colhead{T
- JD$_{0}$} & \colhead{$\mathrm{Flux_{prim}}$} &
\colhead{$\mathrm{Flux_{prim}}$}  &\colhead{$\mathrm{Flux_{comp}}$} & \colhead{$\mathrm{Flux_{comp}}$}\\
  \colhead{}&\colhead{($^{\circ}$)}&\colhead{Position}&\colhead{[hour]}&\colhead{[$\mathrm{e^{-}/s}$]}
  &\colhead{normalized}&\colhead{[$\mathrm{e^{-}/s}$]}&\colhead{normalized}
}

\startdata
       &           &         &       &          &        &        &        \\
       \multirow{11}{*}{1} &  \multirow{11}{*}{202}
                   &       1 & 0.000 & 76400.23 & 0.9842 & 197.47 & 1.0255 \\
      &          &       1 & 0.030 & 77067.92 & 0.9928 & 194.68 & 1.0110 \\
      &          &       1 & 0.059 & 77335.85 & 0.9963 & 194.17 & 1.0084 \\
      &         &       2 & 0.192 & 77461.98 & 0.9907 & 192.72 & 0.9860 \\
      &         &       2 & 0.221 & 77498.89 & 0.9912 & 197.69 & 1.0114 \\
      &         &       2 & 0.251 & 77807.67 & 0.9951 & 194.43 & 0.9947 \\
      &          &       3 & 0.354 & 76836.70 & 0.9896 & 189.01 & 1.0205 \\
      &          &       3 & 0.384 & 77254.05 & 0.9950 & 188.09 & 1.0155 \\
      &          &       3 & 0.413 & 77474.72 & 0.9978 & 184.35 & 0.9953 \\
      &          &       4 & 0.546 & 77188.85 & 0.9915 & 182.92 & 0.9843 \\
      &          &       4 & 0.575 & 77384.02 & 0.9940 & 190.81 &1.0267 \\
      \hline
       &           &         &       &          &        &        &        \\
       \multirow{11}{*}{2} &  \multirow{11}{*}{227}
                 &       1 & 1.594 & 77839.67 & 0.9990 & 194.62 & 1.0401 \\
     &          &       1 & 1.624 & 77952.59 & 1.0005 & 191.95 & 1.0258 \\
     &          &       1 & 1.653 & 77874.77 & 0.9995 & 188.82 & 1.0091 \\
     &          &       2 & 1.786 & 77221.62 & 0.9920 & 192.39 & 0.9974 \\
     &          &       2 & 1.815 & 77989.91 & 1.0019 & 193.38 & 1.0025 \\
     &          &       2 & 1.845 & 77810.97 & 0.9996 & 200.23 & 1.0380 \\
     &          &       3 & 1.948 & 78412.56 & 0.9989 & 199.15 & 1.0223 \\
     &          &       3 & 1.978 & 78506.68 & 1.0001 & 196.32 & 1.0077 \\
     &          &       3 & 2.007 & 78569.23 & 1.0009 & 196.42 & 1.0083 \\
     &          &       4 & 2.139 & 77237.39 & 0.9956 & 202.16 & 1.0228 \\
     &          &       4 & 2.169 & 77530.01 & 0.9994 & 203.36 &
     1.0289 \\
           \hline
       &           &         &       &          &        &        &        \\
       \multirow{12}{*}{3} &  \multirow{12}{*}{202}
                &       1 & 3.188 & 77956.86 & 1.0043 & 191.32 & 0.9936 \\
      &        &       1 & 3.218 & 78179.49 & 1.0072 & 191.70 & 0.9956 \\
      &        &       1 & 3.247 & 78213.98 & 1.0076 & 195.27 & 1.0141 \\
      &        &       2 & 3.379 & 78591.12 & 1.0051 & 203.01 & 1.0386 \\
      &        &       2 & 3.409 & 78608.24 & 1.0054 & 197.20 & 1.0089 \\
      &        &       2 & 3.439 & 78847.12 & 1.0084 & 195.33 & 0.9993 \\
      &        &       3 & 3.542 & 77870.59 & 1.0029 & 184.67 & 0.9971 \\
      &        &       3 & 3.571 & 78204.56 & 1.0072 & 182.92 & 0.9876 \\
      &        &       3 & 3.601 & 78190.32 & 1.0070 & 185.51 & 1.0016 \\
      &        &       4 & 3.733 & 77932.39 & 1.0010 & 187.29 & 1.0078 \\
      &        &       4 & 3.763 & 78139.97 & 1.0037 & 186.62 & 1.0041 \\
      &        &       4 & 3.793 & 78320.57 & 1.0060 & 186.88 & 1.0056
      \\
           \hline
       &           &         &       &          &        &        &        \\
       \multirow{12}{*}{4} &  \multirow{12}{*}{227}
                  &       1 & 4.782 & 77939.05 & 1.0003 & 185.86 & 0.9933 \\
      &          &      1 & 4.811 & 78127.91 & 1.0027 & 189.97 & 1.0153 \\
      &          &       1 & 4.841 & 78240.98 & 1.0042 & 184.40 & 0.9855 \\
      &          &       2 & 4.973 & 77857.88 & 1.0002 & 184.41 & 0.9560 \\
      &          &       2 & 5.003 & 78020.08 & 1.0023 & 190.59 & 0.9880 \\
      &          &       2 & 5.033 & 78140.80 & 1.0038 & 198.42 & 1.0286 \\
      &          &       3 & 5.136 & 78268.69 & 0.9971 & 191.85 & 0.9848 \\
      &          &       3 & 5.165 & 78559.50 & 1.0008 & 195.65 & 1.0043 \\
      &          &       3 & 5.195 & 78640.39 & 1.0018 & 196.62 & 1.0093 \\
      &          &       4 & 5.327 & 77340.36 & 0.9969 & 198.37 & 1.0037 \\
      &          &       4 & 5.357 & 77671.12 & 1.0012 & 196.15 & 0.9924 \\
      &          &       4 & 5.387 & 77871.34 & 1.0038 & 194.04 &
      0.9818 \\
           \hline
       &           &         &       &          &        &        &        \\
       \multirow{12}{*}{5} &  \multirow{12}{*}{202}
                &       1 & 6.391 & 77600.12 & 0.9997 & 189.87 & 0.9861 \\
      &          &       1 & 6.421 & 77859.67 & 1.0030 & 188.72 & 0.9800 \\
      &          &       1 & 6.450 & 77996.08 & 1.0048 & 189.82 & 0.9858 \\
      &          &       2 & 6.582 & 78167.24 & 0.9997 & 195.86 & 1.0020 \\
      &          &       2 & 6.612 & 78393.06 & 1.0026 & 193.02 & 0.9875 \\
      &          &       2 & 6.642 & 78320.96 & 1.0017 & 189.91 & 0.9716 \\
      &          &       3 & 6.745 & 77489.71 & 0.9980 & 184.65 & 0.9970 \\
      &          &       3 & 6.774 & 77745.72 & 1.0013 & 183.84 & 0.9926 \\
      &          &       3 & 6.804 & 77746.48 & 1.0013 & 183.86 & 0.9927 \\
      &          &       4 & 6.936 & 77799.03 & 0.9993 & 181.59 & 0.9771 \\
      &          &       4 & 6.966 & 77910.98 & 1.0007 & 185.86 & 1.0001 \\
      &          &       4 & 6.996 & 78149.82 & 1.0038 & 184.81 &
     0.9944 \\
                \hline
       &           &         &       &          &        &        &        \\
       \multirow{12}{*}{6} &  \multirow{12}{*}{227}
           &       1 & 8.063 & 77432.59 & 0.9938 & 185.75 & 0.9927 \\
 &          &       1 & 8.092 & 77916.03 & 1.0000 & 182.53 & 0.9755 \\
 &          &       1 & 8.122 & 77916.40 & 1.0000 & 180.14 & 0.9627 \\
 &          &       2 & 8.254 & 77800.14 & 0.9995 & 188.38 & 0.9766 \\
 &          &       2 & 8.284 & 77805.98 & 0.9995 & 193.43 & 1.0028 \\
 &          &       2 & 8.314 & 77923.23 & 1.0011 & 194.87 & 1.0102 \\
 &          &       3 & 8.417 & 78389.06 & 0.9986 & 194.13 & 0.9965 \\
 &          &       3 & 8.446 & 78630.61 & 1.0017 & 192.12 & 0.9862 \\
 &          &       3 & 8.476 & 78515.42 & 1.0002 & 191.06 & 0.9807 \\
 &          &       4 & 8.608 & 77414.03 & 0.9979 & 198.44 & 1.0040 \\
 &          &       4 & 8.638 & 77724.12 & 1.0019 & 194.72 & 0.9852 \\
 &          &       4 & 8.668 & 77842.54 & 1.0034 & 193.93 &
     0.9812 \\
     \enddata
\end{deluxetable}

\begin{deluxetable}{cccccccc}
\tabletypesize{\scriptsize}
\tablecaption{F160W Photometry Results\label{tab:F160W}}
\tablewidth{0pt}
\tablehead{
\colhead{Orbit} & \colhead{Pos. Angle} & \colhead{Dither} & \colhead{T
- JD$_{0}$} & \colhead{$\mathrm{Flux_{prim}}$} &
\colhead{$\mathrm{Flux_{prim}}$}  &\colhead{$\mathrm{Flux_{comp}}$} & \colhead{$\mathrm{Flux_{comp}}$}\\
  \colhead{}&\colhead{($^{\circ}$)}&\colhead{Position}&\colhead{[hour]}&\colhead{[$\mathrm{e^{-}/s}$]}
  &\colhead{normalized}&\colhead{[$\mathrm{e^{-}/s}$]}&\colhead{normalized}
}

\startdata
       &           &         &       &          &        &        &
       \\
              \multirow{10}{*}{1} &  \multirow{10}{*}{202}
        &       1 & 0.091 & 63247.39 & 0.9949 & 309.55 & 0.9789 \\
&        &       1 & 0.121 & 63231.17 & 0.9947 & 320.57 & 1.0138 \\
&        &       1 & 0.150 & 63234.83 & 0.9947 & 310.39 & 0.9816 \\
&        &       2 & 0.282 & 62880.74 & 0.9899 & 314.73 & 0.9965 \\
&        &       2 & 0.312 & 63170.96 & 0.9945 & 319.87 & 1.0127 \\
&        &       3 & 0.445 & 63482.02 & 0.9909 & 312.57 & 1.0003 \\
&        &       3 & 0.474 & 63876.58 & 0.9970 & 314.81 & 1.0075 \\
&        &       3 & 0.504 & 63868.25 & 0.9969 & 316.42 & 1.0126 \\
&        &       4 & 0.607 & 63268.09 & 0.9911 & 305.45 & 0.9924 \\
&        &       4 & 0.636 & 63637.60 & 0.9969 & 310.09 & 1.0074 \\
\hline
       &           &         &       &          &        &        &
       \\
              \multirow{10}{*}{2} &  \multirow{10}{*}{227}
        &       1 & 1.685 & 63800.82 & 1.0014 & 317.96 & 1.0094 \\
&        &       1 & 1.714 & 63646.56 & 0.9990 & 320.24 & 1.0167 \\
&        &       1 & 1.744 & 63712.52 & 1.0001 & 318.32 & 1.0106 \\
&        &       2 & 1.876 & 62867.13 & 0.9951 & 327.52 & 1.0173 \\
&        &       2 & 1.906 & 63215.33 & 1.0007 & 322.00 & 1.0001 \\
&        &       3 & 2.039 & 63472.43 & 0.9940 & 312.25 & 1.0059 \\
&        &       3 & 2.068 & 64211.58 & 1.0056 & 314.82 & 1.0142 \\
&        &       3 & 2.098 & 64018.46 & 1.0025 & 307.83 & 0.9917 \\
&        &       4 & 2.201 & 63356.87 & 0.9965 & 315.89 & 1.0031 \\
&        &       4 & 2.230 & 63173.89 & 0.9937 & 318.91 & 1.0127 \\
\hline
       &           &         &       &          &        &        &
       \\
              \multirow{11}{*}{3} &  \multirow{11}{*}{202}
        &       1 & 3.279 & 63508.62 & 0.9991 & 321.56 & 1.0169 \\
&        &       1 & 3.308 & 63639.15 & 1.0011 & 321.16 & 1.0156 \\
&        &       1 & 3.338 & 63817.51 & 1.0039 & 318.01 & 1.0057 \\
&        &       2 & 3.470 & 63713.27 & 1.0030 & 317.28 & 1.0046 \\
&        &       2 & 3.500 & 63948.17 & 1.0067 & 318.58 & 1.0087 \\
&        &       3 & 3.632 & 64133.71 & 1.0010 & 318.27 & 1.0185 \\
&        &       3 & 3.662 & 64148.99 & 1.0013 & 316.10 & 1.0116 \\
&        &       3 & 3.692 & 64378.74 & 1.0049 & 310.91 & 0.9950 \\
&        &       4 & 3.824 & 64067.16 & 1.0036 & 300.93 & 0.9777 \\
&        &       4 & 3.854 & 64067.97 & 1.0036 & 308.13 & 1.0011 \\
&        &       4 & 3.884 & 63910.80 & 1.0011 & 314.97 & 1.0233 \\
\hline
       &           &         &       &          &        &        &
       \\
              \multirow{11}{*}{4} &  \multirow{11}{*}{227}
        &       1 & 4.872 & 63914.07 & 1.0032 & 311.23 & 0.9880 \\
&        &       1 & 4.902 & 63884.46 & 1.0028 & 315.39 & 1.0013 \\
&        &       1 & 4.932 & 63920.45 & 1.0033 & 314.80 & 0.9994 \\
&        &       2 & 5.064 & 63364.21 & 1.0030 & 321.69 & 0.9992 \\
&        &       2 & 5.094 & 62992.57 & 0.9971 & 320.14 & 0.9944 \\
&        &       3 & 5.226 & 63789.12 & 0.9990 & 310.93 & 1.0017 \\
&        &       3 & 5.256 & 64018.91 & 1.0025 & 312.98 & 1.0083 \\
&        &       3 & 5.286 & 63849.69 & 0.9999 & 309.99 & 0.9986 \\
&        &       4 & 5.418 & 63722.64 & 1.0023 & 316.11 & 1.0038 \\
&        &       4 & 5.448 & 63707.71 & 1.0021 & 312.25 & 0.9915 \\
&        &       4 & 5.477 & 63813.00 & 1.0037 & 314.67 & 0.9992 \\
\hline
       &           &         &       &          &        &        &
       \\
              \multirow{11}{*}{5} &  \multirow{11}{*}{202}
        &       1 & 6.482 & 63771.50 & 1.0032 & 306.82 & 0.9703 \\
&        &       1 & 6.511 & 63910.88 & 1.0054 & 316.82 & 1.0019 \\
&        &       1 & 6.541 & 63759.91 & 1.0030 & 321.02 & 1.0152 \\
&        &       2 & 6.673 & 63587.56 & 1.0011 & 315.41 & 0.9986 \\
&        &       2 & 6.703 & 63817.91 & 1.0047 & 309.17 & 0.9789 \\
&        &       3 & 6.836 & 64135.94 & 1.0011 & 303.13 & 0.9701 \\
&        &       3 & 6.865 & 64302.83 & 1.0037 & 306.83 & 0.9819 \\
&        &       3 & 6.895 & 64273.25 & 1.0032 & 313.31 & 1.0027 \\
&        &       4 & 7.027 & 64005.66 & 1.0026 & 304.82 & 0.9903 \\
&        &       4 & 7.057 & 63728.45 & 0.9983 & 307.86 & 1.0002 \\
&        &       4 & 7.087 & 64020.34 & 1.0029 & 310.13 & 1.0076 \\
\hline
       &           &         &       &          &        &        &
       \\
              \multirow{11}{*}{6} &  \multirow{11}{*}{227}
        &       1 & 8.154 & 63248.40 & 0.9928 & 314.61 & 0.9988 \\
&        &       1 & 8.183 & 63573.69 & 0.9979 & 309.55 & 0.9827 \\
&        &       1 & 8.213 & 63681.11 & 0.9996 & 312.85 & 0.9932 \\
&        &       2 & 8.345 & 63367.46 & 1.0031 & 317.65 & 0.9866 \\
&        &       2 & 8.375 & 63238.70 & 1.0010 & 322.75 & 1.0025 \\
&        &       3 & 8.508 & 63746.16 & 0.9983 & 306.08 & 0.9861 \\
&        &       3 & 8.537 & 63724.29 & 0.9979 & 309.73 & 0.9978 \\
&        &       3 & 8.567 & 63874.72 & 1.0003 & 309.08 & 0.9957 \\
&        &       4 & 8.699 & 63614.87 & 1.0006 & 309.17 & 0.9817 \\
&        &       4 & 8.729 & 63530.79 & 0.9993 & 320.22 & 1.0168 \\
&        &       4 & 8.759 & 63696.84 & 1.0019 & 312.14 & 0.9912 \\
\enddata
\end{deluxetable}

\end{document}